# An extensive thermal conductivity measurement method based on atomic force microscopy

T. Serkan Kasırga[1,2*], Berke Köker[3]
[1] Bilkent University UNAM – National Nanotechnology Research Center, Bilkent, Ankara 06800 Türkiye
[2] Institute of Material Science and Nanotechnology, Bilkent University, Ankara 06800 Türkiye
[3] Department of Physics, Bilkent University, Bilkent, Ankara 06800 Tükiye
**\*Corresponding author:** kasirga@unam.bilkent.edu.tr

**Abstract**
Heat transport in low-dimensional solids can significantly differ from their bulk counterpart due to various size-related effects. This offers rich heat transport phenomena to emerge. However, finding an appropriate thermometry method for thermal conductivity measurements at the reduced size and dimensionality of the samples is a challenge. Here, we propose and study the feasibility of a nanoscale resolution thermal conductivity measurement method based on bolometric thermometry implemented on an atomic force microscopy (AFM). The local heat exchange between the AFM tip and the sample occurs at a suspended section of the sample, and thermal modeling of the measured electrical resistance change resulting from the bolometric effect provides a unique value for thermal conductivity. As we illustrate via thermal simulations, the proposed method can measure thermal conductivity with thermal disturbance to the sample in as little as 0.2 K at ~20 nm lateral resolution. Our in-depth analysis shows the feasibility and extensive applicability of the proposed AFM-based bolometric thermometry method on low-dimensional materials both in diffusive and ballistic heat transport regimes from cryogenic to above-room temperature. Consequently, the proposed method can lead to a deeper experimental understanding of fundamental questions in nanoscale and low-dimensional heat transport phenomena in many different material classes, as well as Fourier and non-Fourier heat transfer regimes.

**Main Text**

**Introduction**
Thermal gradients are ubiquitous in solid-state devices as they rarely operate under thermal equilibrium. The resultant heat flow from the hotter parts of the material to the colder parts can be quantified by thermal conductivity ($\kappa$), a critical parameter for effective thermal management in electronic, optical, and quantum devices. Heat transport is sensitive to all energy-carrying degrees of freedom in a solid[1]. As a result, thermal conductivity can be used as a selective probe of solid-state phenomena. By utilizing thermal conductivity measurements, a deeper understanding of strongly correlated, classical, and quantum phenomena can be achieved.

From a broad perspective, the open challenges that systematic thermal conductivity measurements can be instrumental in tackling can be listed as follows: **(1)** Non-Fourier heat transport regimes haven't been explored coherently and reproducibly. Apart from the classical size effects on nanowires and thin films, investigation of other heat transport regimes such as Ioffe-Regel, hydrodynamic, Anderson localized, and coherent phonon regimes have been largely unexplored and have potential technological significance[2]. Moreover, questions like why carbon nanotubes and graphene exhibit exceptionally high thermal conductivity despite the Casimir-Knudsen effect lie as open questions. **(2)** The extent of the breakdown of the Wiedemann-Franz (WF) law[3] is not well-known. For weakly interacting electronic systems, the WF law provides a robust empirical relation between the thermal conductivity $\kappa$ and the electrical conductivity $\sigma$ of a material. The breakdown of WF law typically indicates a departure from the weakly interacting fermion picture, namely the Fermi-liquid (FL) model[4]. Moreover, for materials where WF is not valid, it is unclear whether there is another relation between other material parameters and thermal conductivity. In the past two decades, non-trivial violations of WF law have been demonstrated in various strongly interacting systems[5–9]. Heat carriers beyond the quasiparticles of the FL model, like spinons and holons in strictly one-dimensional systems that obey Tomonaga-Luttinger liquid (TLL) states[8,10,11] exhibit strong departures from the WF law. At the verge of phase transitions and in low dimensions[12], the breakdown of WF can be used as a probe to elucidate the underlying physical phenomena. **(3)** Heat transport measurements can be pivotal in identifying exotic quantum states such as quantum spin liquids[13] and can be used in detecting topological quantum materials such as Weyl semimetals and topological insulators via the Nernst (or anomalous Nernst) effect[14–17]. However, a systematic approach in low-dimensional quantum materials is missing in the literature. In particular, temperature, anisotropy, and magnetic field-dependent thermal conductivity measurements can open a powerful window to the exotic quantum states by providing a phase diagram[18].

**Table 1.** Qualitative comparison of thermometry methods commonly used for micro and nano-scale materials on three major criteria. The last row provides an expected parameter range for the method proposed here. The color scheme illustrates the relative advantage of the method on a given parameter. Green, yellow, and red indicate the most, medium, and least advantageous conditions on the given parameter, respectively.

| Method | (1) Invasiveness | (2) Heater / Thermometer size | (3) Extrinsic parameter variability |
|---|---|---|---|
| Microbridge thermometry | Minimal | Depends on the sample properties | Wide |
| Raman thermometry | Large | Diffraction Limited | Limited |
| 3ω method | Large | Limited by the sample thickness | Limited |
| Scanning thermal microscopy[19,20] | Minimal | Below 100 nm | Limited |
| Time/frequency domain thermoreflectance | Limited | Limited | Limited |
| Transient thermal gratings[21,22] | Moderate | Limited by grating wavelength | Wide |
| SQUID on tip[23,24] | Minimal | Below 100 nm / Not a heater | Restricted $T_C$ of SQUID |
| *Mechanical bolometric thermometry* | Minimal / Tapping or non-contact heating | Variable / 20-200 nm | 1.6-500 K, 9 Tesla, large selection of material classes |

To measure the thermal conductivity of a material, a precise determination of the temperatures at two different positions of the specimen is required. Despite the conceptual and practical simplicity of measuring the temperature at the macroscopic scale, thermometry at the nano and micro scale to extract thermal conductivity is challenging. This is mostly due to the involvement of the classical and quantum size effects on heat transport, as well as challenges involving the fabrication of a non-invasive thermometer. Although there is rapid progress in the field of heat transport at the nanoscale, available thermometry methods are not practically applicable to a wide variety of solids[25,26]. **Table 1** provides a qualitative comparison between commonly used thermometry methods to study heat transport and thermal conductivity at micro and nano-scale materials. The qualities referred to in the table are determined by the parameters mentioned in the literature

To answer the outstanding open questions outlined in the previous paragraphs, a widely applicable tool with minimal thermal disturbance to the sample and independent of the materials' type and extrinsic properties is required. Recently, we demonstrated that bolometric thermometry could be implemented using optical excitation as a heat source, such that the material under study is used as an optical bolometer to extract the local temperature increase[25,27]. We demonstrated that very small temperature-induced resistivity changes (~100 ppm) could be measured by suitably used lock-in amplifiers, and thus, thermal profiles can be extracted via this so-called optical bolometric thermometry (OBT). This simple idea enabled fast, sensitive, and accurate thermometry that applies to quantifying $\kappa$ in one and two-dimensional (1&2D) metals. In another study[28], we realized that a hot tip instead of light might be used to induce a bolometric and thermoelectric response, which, in principle, removes the limitations of the OBT and can be used on a much wider range of materials.

Here, we propose and numerically show that an atomic force microscope (AFM) tip hotter than the sample can be used to achieve the desired thermal conductivity measurement criteria for exploring the nanoscale heat transport phenomena. As the bolometric response is induced by the mechanical contact of the AFM tip at a temperature different than that of the sample, we will refer to this method as Mechanical Bolometric Thermometry (MBT). The basic operational principle of the proposed method could provide a universally applicable thermal conductivity measurement method for metallic and semiconducting materials. We used commercial finite element analysis software (COMSOL Multiphysics®) along with experimental material parameters to calculate the measurement performance of MBT. Moreover, we proposed and discussed experimental configurations that can be implemented inside or outside a cryostat. As a result, low temperature and quantum phenomena can be studied via the proposed method both at the diffusive and ballistic heat transport regimes. In this paper, we provide a systematic approach to elucidate the method, exhibit its fundamental principles and experimental implementation, and discuss its limitations.

**Fundamental principles of mechanical bolometric thermometry**
Although MBT can work for one-, two-, and three-dimensional geometries, we will focus on its applicability to two-dimensional materials. There are three essential elements of tip-based mechanical bolometric thermometry: (1) AFM tip with a temperature $T_{AFM}$, different than the ambient temperature, $T_{Amb}$. We assume that the sample is in thermal equilibrium with the ambient, (2) the sample under test is suspended over a circular hole or a trench, such that it is partially isolated from the substrate, and (3) the sample is electrically contacted. The bolometric thermometry relies on the temperature change at the suspended part of the sample. In the case of MBT, the temperature change is caused by the intermittent contact of the AFM tip with the sample. Suspending the sample is important to eliminate any complications that may arise from thermal boundary conductance between the material and the substrate; however, stacked structures can also be studied by appropriate modeling.

The aim of the thermometry is to find the temperature change, $\delta T$, at the point of intermittent contact with the AFM tip. Due to the thermal boundary losses and radiation losses, any thermometer on the AFM tip would provide a temperature reading that is vastly different from that of the sample at the point of contact. Moreover, there is no experimentally demonstrated way to accurately measure the temperature of the sample at the point of contact, $T^*_{AFM}$. This is the major challenge in using scanning thermal microscopy (SThM), a variant of AFM with a thermometer and heater embedded in it, as a tool for thermal transport measurements. In MBT, we focus on the temperature change at the point of contact, which can be written as $\delta T = |T^*_{AFM} - T_{Amb}|$. This change can be measured and quantified by measuring the electrical resistance change of the sample, with thermal conductivity being a fitting parameter, which we will detail now.

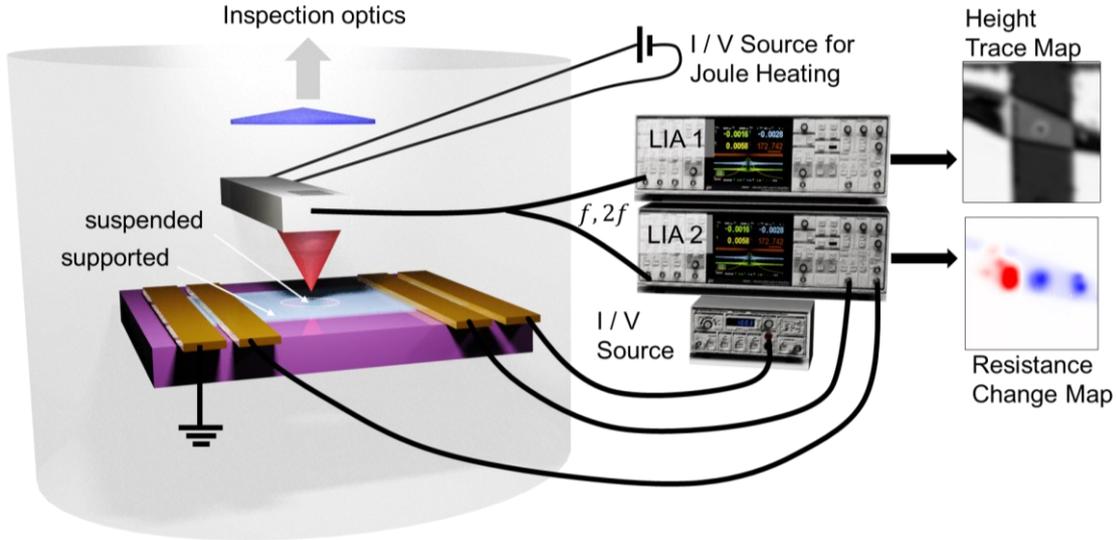

**Figure 1| Experimental schematic of MBT.** A four-terminal device of a two-dimensional sample suspended over a circular hole is depicted with a heated AFM tip scanning over the sample. Two lock-in amplifiers are used to extract the height trace map and resistance change map. Current/Voltage (I/V) bias is applied through a source connected to the outer contacts. The entire setup can be implemented inside a controlled environment for low-temperature measurements.

Depending on the material type, whether it is metallic or semiconducting, the electrical resistivity of a material, $\rho(T)$, depends on its temperature. Moreover, this dependence can be extended to low-temperature phenomena such as charge density wave transitions, Kondo effect, and Fermi-Liquid transitions. The list of phenomena with strong temperature-dependent resistivity can be further expanded. For the case of metals, in the linear approximation, resistivity typically follows $\rho(T) = \rho_0[1 + \alpha(T - T_0)]$ relation whereas when the Kondo effect dominates, it can be expressed $\rho(T) = \rho_0 + aT^2 + c_m \ln \mu/T + bT^5$. Here, $aT^2$ is the Fermi liquid contribution, $bT^5$ is from the lattice vibrations, and $a, b, c_m, \mu$ are constants.

When the AFM tip is in intermittent contact with the sample surface, the resistivity of the sample can be defined as $\rho[\delta T(x, y)]$. Here, $\delta T(x, y)$ denotes the temperature distribution over the suspended (and also supported, but its effect will be negligible as we will discuss later) part of the sample. As a result, when the AFM tip is on the suspended part of the material, there will be a temperature distribution determined by the thermal conductivity $\kappa$ or the material. The governing equation for will be:

$$r < a \qquad \kappa \frac{1}{r} \frac{d}{dr}\left[r \frac{dT_1(r)}{dr}\right] + \frac{Q}{t} e^{-r^2/r_0^2} = 0 \qquad \text{(Eq. 1)}$$

where, $T_1(r)$ is the radial temperature distribution over the circular hole of the radius $a$, $Q$ is the power delivered/extracted from the sample, $r_o$ is the effective thermal radius of the AFM tip and $t$ is the thickness of the sample. Another equation is required for $r \geq a$ region of the sample along with the boundary conditions. Further details of the heat equations at $r < a$ and $r \geq a$ regions are discussed elsewhere[25,27]. So, if $\rho[\delta T(x, y)]$ is measured, then by using $\kappa$ as a fitting parameter and with the known $\rho(T)$ relation and $Q$, $\delta T(x,y)$ can be extracted. The fitting provides a single $\kappa$ value to fit the $\delta T(x,y)$ for the measured resistance change value[27].

**Experimental Implementation of MBT**

One of the most peculiar aspects of the OBT and MBT methods is that measuring thermal conductivity becomes a straightforward electrical measurement in the presence of a local heat source followed by a parametric fitting. **Figure 1** depicts the proposed experimental configuration of the MBT. An SThM configuration would give more control over the measurements and would allow quantitative determination of the thermal power, $Q$ in **Eq.1**, delivered to the sample for a more precise extraction of $\kappa$. However, a sample heater or cooler with a regular AFM setup would also work with reduced precision, as $Q$ can be predicted based on the heat transfer across the sample and the tip. For the sake of brevity, we discuss how $Q$ can be extracted in SThM configuration in the Supporting Information[29–31].

A bias source and a lock-in amplifier would be connected to the sample for AC resistance measurement. As a consequence, very high signal-to-noise ratios can be achieved. The modulation frequency of the AC measurement would be the same as the AFM tip modulation frequency in the tapping mode. Thus, the measurement can become insensitive to environmental thermal fluctuations. Moreover, the measurement sensitivity can be further improved by implementing a four-terminal electrical configuration, as depicted in **Figure 1**. Here, the electrical bias can be applied through the outer contacts, and the resistance change can be measured from the inner contacts. The piezo-scanner position would be fed to a computer along with the output of the lock-in amplifier to plot the x-y plot of the tip-induced resistance change on the electrically biased sample. Scanning is not mandatory; however, it would significantly reduce the positioning errors of the tip to the center of the suspended section of the sample and provide geometric parameters of the sample, which is needed for extracting the electrical resistivity. Finally, the AFM tip would also provide height trace information during the scan. As a result, the spatial distribution of the resistance change can be correlated to the height trace map. On a separate measurement, the temperature-dependent resistivity of the sample must be extracted for thermal conductivity fitting. This can be either performed using the Peltier plate with a temperature range limited to that of the measurement, or a larger temperature range measurements can be performed within a cryostat. With the information from the AFM height trace, the temperature-dependent sample resistivity, $\rho(T)$, can be extracted.

**Temperature change induced by the AFM tip**

To elucidate the feasibility of the proposed method, we performed analytical and finite element analysis modeling based on real material parameters and calculated the performance of the MBT using commercial FAE software, COMSOL Multiphysics®. The first question regarding the feasibility of the method is whether the AFM tip can induce temperature change large enough on the sample to induce a measurable resistivity change or not. AFM tip has a small effective heat exchange radius (~20 nm). As a result, the heat exchange area of the sample would be much smaller than that of the previously experimentally demonstrated OBT method[32]. To compare the temperature change induced by the AFM tip on the sample between MBT and OBT, we performed FEA simulations. **Figure 2a** and **b** show the simulated temperature change map of a 30 nm thick 2H-TaS$_2$ with 175 Ω resistance suspended over a hole of 2 μm radius for OBT and MBT, respectively. We used a Gaussian beam profile with 400 nm full width at half maximum for the OBT model and a 20 nm-radius AFM tip with a Gaussian heating profile for the MBT model. Constant power is delivered to the sample in both cases. FEA results show that in MBT, despite the sharp drop in temperature towards the edges of the suspended region, there is a large enough temperature change to induce a bolometric response.

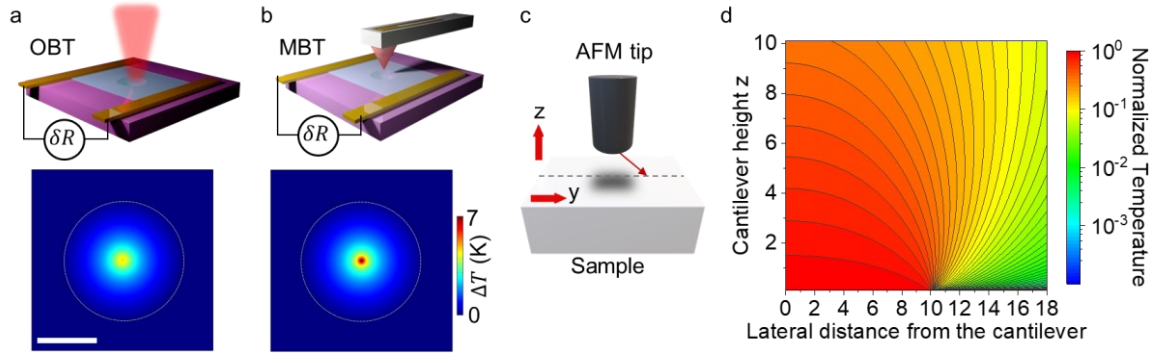

**Figure 2 | OBT vs. MBT and the heating by the AFM tip.** Schematics of the OBT and MBT measurement configurations and the thermal distribution maps when the heat source is at the center are given in **a** and **b**, respectively. White dashed circles mark the suspended region of the sample. The scale bar represents 2 μm. **c** Schematic of the model used for calculating the thermal distribution on the sample versus the cantilever height from the surface. AFM tip is modeled as a cylinder with a diameter of $D$ and all the molecules scatter from the bottom surface of the tip. The lateral distance from the tip is taken to be the y-axis. **d** Normalized temperature of the sample (i.e., the temperature of the cantilever is set to 1) versus cantilever height from the surface and the lateral distance from the center of the cantilever is shown. The sample temperature decreases by an order of magnitude when the tip is $D/2$ away from the sample surface.

A tapping AFM tip can be modeled to determine the spatial resolution of the MBT and how tapping the tip would modulate the sample temperature. Here, in our model, we assume that the measurement takes place in a gaseous environment, either in the ambient or in the presence of He[4] as a heat exchange gas. Moreover, thermal radiation from or to the tip would contribute to the heat transfer. When the tip is in direct contact with the sample, then the heat will be transferred predominantly through mechanical contact. We neglect the contributions from the moisture meniscus around the tip as this will be minimal in moisture-controlled environments. Thus, for the sake of simplicity, we assume that while the apex of the tip is in contact with the sample, we can assume that the sample temperature at the disc of contact will be $\delta T + T_{Amb} = T^*_{AFM}$. When the tip is lifted, the heat transfer kernel $H(x, y)$ can be described to model the heat transfer. The apex of the AFM tip can be modeled as a disc of diameter $D$. We assume each exchange gas molecule carries excess energy proportional to $T_{AFM}$ and they scatter uniformly from the apex and, a uniform temperature distribution at the cantilever apex. Based on these assumptions, we can write the ratio of heat exchange gas reaching from the hot tip to the surface as the ratio of a small solid angel $\Delta\Omega$ and sample surface area $\Delta S$ as $\frac{\Delta\Omega}{\Delta S} = z(z^2 + r^2)^{-3/2}$ and

$$H(x,y) = \frac{1}{2\pi}\int_0^{D/2} d\rho\rho \int_0^{2\pi} d\phi z(z^2 + (x - \rho\cos\phi)^2 + (y - \rho\sin\phi)^2)^{-3/2} \quad \text{(Eq. 2)}$$

as an integral form in azimuthal coordinates. Accordingly, temperature change on the sample surface due to the AFM tip at height $z > 0$ can be written as:

$$\delta T(x,y) = \frac{4\delta T}{\pi D^2} \int dx' \int dy' H(x - x', y - y') \quad \text{(Eq. 3)}$$

**Figure 2c** depicts the model and **Figure 2d** gives the map of the $\delta T(0, y)$ vs $z$. This shows that the largest temperature rise will occur at the intermittent contact, and this will set the resolution of MBT almost as close to the tip apex diameter. As the tip moves up by half a diameter, the sample temperature reduces by an order of magnitude. This also shows that there will be sufficient thermal modulation for lock-in amplification to work as typically amplitude of the AFM tip oscillation ranges between 10-100 nm.

Next, we focused on three different samples representing different types of materials. To explore the effect of the thermal conductivity of the materials on the maximum temperature of the sample, $\delta T$, and the total resistance change of the device, $\delta R$, we plotted the simulated values for $\delta T$ and $\delta R$ versus the fitted thermal conductivity of the materials, given in **Figure 3**. As the materials become more thermally insulating, the local temperature of the sample increases under the constant heat from the excitation source. We would like to reiterate that in experiments, $\delta R$ is measured, and $Q$ can be controlled, thus a single $\kappa$ value for the material can be obtained. To provide a comparison, both OBT and MBT are studied in the simulations since OBT has been experimentally verified in our previous measurements. Simulations are performed under constant power delivery from the tip (or from the optical excitation for the case of OBT) to the sample, and we assumed the heat is delivered from the tip to the sample rather than extracted from the sample. We also calculated the noise floor, as discussed in the following section, to provide a sensitivity floor for the measurements.

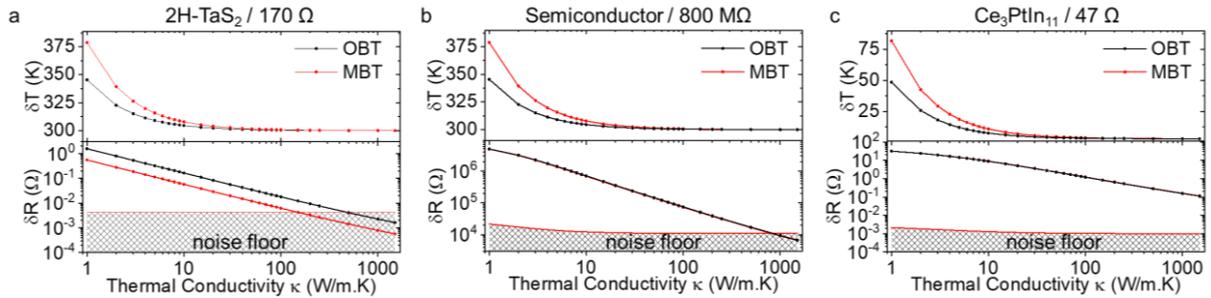

**Figure 3 | Comparison of OBT and MBT for three different materials.** Under constant heat influx through the center of the suspended part of samples, $\delta T$ and $\delta R$ values are plotted against fitted thermal conductivity for **a** 2H-TaS$_2$, **b** a hypothetical semiconductor and, **c** Ce$_3$PtIn$_{11}$ heavy fermion superconductor above $T_c$. The intrinsic noise floor is indicated in all three cases.

First, we simulated 30 nm thick 2H-TaS$_2$ as a layered metallic material with 175 Ω resistance. One of the reasons we focus on 2H-TaS$_2$ is that we, via OBT[25] and recently Liu et al.[26], studied its thermal conductivity. The resistance change for the case of 2H-TaS$_2$ remains above the intrinsic noise floor for a wide range of thermal conductivity values. Next, we focused on a 30 nm thick hypothetical large gap semiconductor device with 800 MΩ resistance. Akin to the previous case, for a very large range of thermal conductivities, MBT could provide a unique $\kappa$ above the noise floor. Finally, we studied a 30 nm thick heavy fermion superconductor Ce$_3$PtIn$_{11}$ at 3 K, above its superconducting phase transition temperature. Resistivity vs. temperature data is based on reference [33]. At 3 K, modeled Ce$_3$PtIn$_{11}$ exhibits 47 Ω resistance, and the resistance change remains well above the noise floor for a wide range of thermal conductivity values. In all cases 10 mV bias is applied to measure the resistance change.

Although we studied three cases based on constant power delivered from the tip, in experiments, the constant tip would be used. As a result, the maximum temperature will be based on the heat exchange across the sample and the tip and the thermal conductivity of the material. Thus, the MBT measurement will yield a single $\delta R$ value. Furthermore, there will be various noise sources in real experiments, as we will discuss in the following section. The constant power approach also illustrates how tip temperature should be manipulated to measure materials with high thermal conductivity and poor temperature-dependent electrical resistivity.

**Measurement sensitivity of MBT**
Various intrinsic and extrinsic factors would limit the measurement sensitivity of MBT. Intrinsic factors that would limit the measurement sensitivity are due to the fundamental noise sources such as the Johnson noise, shot noise and $1/f$ – noise. Extrinsic limitations are due to electrical contact resistance of the samples, poor determination of heat exchange rate $Q$ across the sample and the tip, and measurement errors in material parameters such as resistivity, geometric factors, etc.

At its core, MBT is a very sensitive resistance change measurement method. As a result, it is fundamentally limited by the Johnson, shot and $1/f$ – noise. Johnson noise is defined as $V_J = \sqrt{4k_B T R \Delta f}$. Here, $k_B$, $T$, $R$, $\Delta f$ are the Boltzmann constant, sample temperature, sample resistance, and the measurement bandwidth, respectively. Shot noise, which arises from the discrete nature of the charge carriers, is defined as $I_s = \sqrt{2qI\Delta f}$. Here, $q$ and $I$ are the unit charge and the current, respectively. Finally, $1/f$ – noise could also be taken into consideration. Typical power spectral density follows $S(f) \propto \frac{1}{f^\alpha}$, where $\alpha$ and $f$ are constant and the frequency, respectively. MBT will operate at the modulation frequency of the AFM cantilever, ~10 kHz. The pre-amplifier can be set to narrow the measurement bandwidth, and the lock-in amplifier equivalent noise bandwidth (ENBW) of the low pass filter, which can be adjusted by the time constant, can be set around the modulation frequency to achieve low Johnson noise. By assuming a flat band, a root-mean-square (rms) noise of $V_J(rms) = 0.13\sqrt{R}\sqrt{\Delta f_{ENBW}}$ nV can be achieved at room temperature. Similarly, shot noise can be factored into the noise estimate. $1/f$ – noise is typically much smaller than the other noise factors at ~10 kHz. Thus, we don't take it into account in our calculations. As demonstrated in three different cases in Figure 3, the intrinsic noise floor (by assuming full bandwidth) is several orders of magnitude lower than the measurement sensitivity for the AFM-based method for a wide range of thermal conductivities.

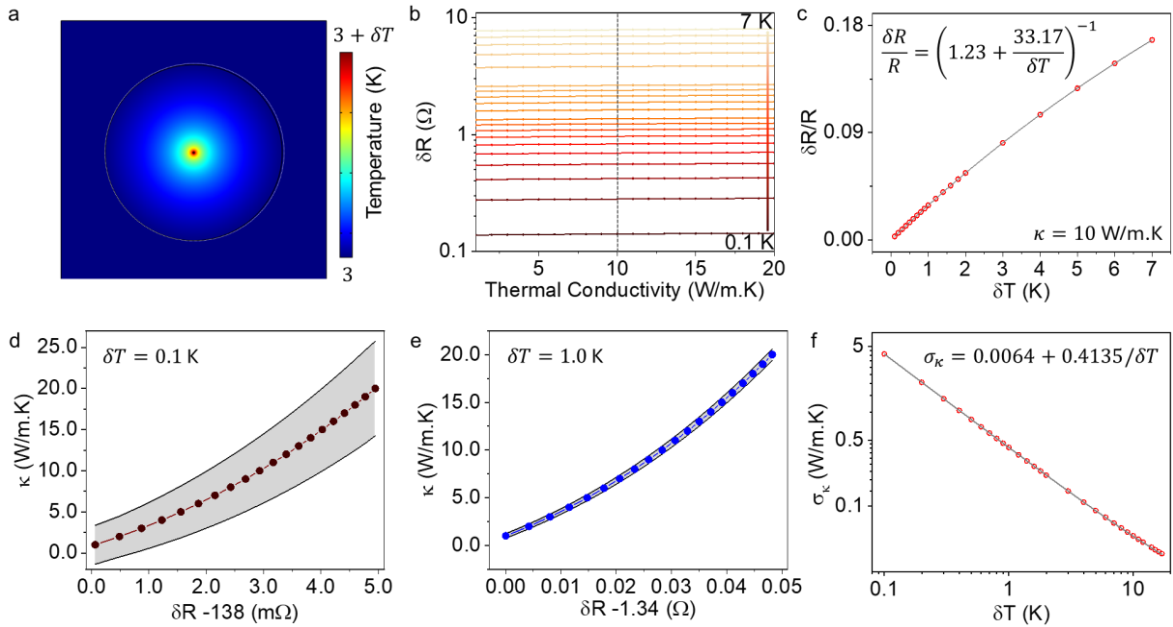

**Figure 4 | Sensitivity of MBT under constant tip temperature. a** Temperature map of a 30 nm thick $Ce_3PtIn_{11}$ sample, when the AFM tip is at the center of the suspended region for kappa, $\kappa = 10$ W/m.K . White dashed circle has a diameter of 4 µm and represents the suspended region of the sample. **b** Resistivity change of the sample for various thermal conductivity values of the sample at different $\delta T$ ranging from 0.1 to 7 K. The Dashed line indicates the $\delta R$ at 10 W/m.K **c** $\delta R/R$ versus $\delta T$ at $\kappa = 10$ W/m.K graph shows the relative change of resistivity for different $\delta T$ values. At above 0.2 K temperature induced by the tip on the sample, measurement sensitivity exceeds the intrinsic noise levels, and measurement sensitivity improves significantly at higher $\delta T$ values. The solid line is the fit given by the equation inset. **d** The thermal conductivity measurement sensitivity window for $\delta T = 0.1$ K is shown. The filled circles represent the indicated $\kappa$, and the grey region shows the error range due to the intrinsic noise sources. **e** Error in determining $\kappa$ decreases significantly when the temperature increase on the sample is 1 K. **f** Graph shows the change or error in determining $\kappa$ as a function of $\delta T$ for $\kappa = 10$ W/m.K. The fit

shows that the error exponentially decreases with the increasing tip-induced sample temperature.

Unlike intrinsic factors, extrinsic factors that may contribute to measurement sensitivity can be eliminated at the cost of increased measurement complexity. For instance, the effect of electrical contact resistance can be eliminated by implementing a four-terminal measurement scheme[34]. As mentioned earlier, $Q$, power transfer across the sample, and the tip can be quantified by using an SThM tip. Also, performing the measurements in an ambient controlled AFM, such as one incorporated inside a cryostat, would limit the convective heat transfer from the sample to the tip and tip to the environment. As a result, better precision can be achieved.

To have a more detailed understanding of how intrinsic noise affects the measurement sensitivity, we performed a further analysis of the results, as shown in detail in **Figure 4**. Again, we used the modeled 30 nm thick $Ce_3PtIn_{11}$ at 3 K that exhibits 47 Ω resistance for the devices that are suspended over a hole of 2 µm radius, using Fourier heat transport. AFM tip is assumed to induce a constant temperature increase upon contact, denoted by $\delta T$. 10 mV electrical bias is applied across the contacts. **Figure 4a** shows the thermal map of the sample for an assumed thermal conductivity of $\kappa = 10$ W/m.K. Resistance change of the sample depends on the thermal conductivity of the material as well as the temperature increase induced by the $\delta T$. **Figure 4b** shows the change of $\delta R$ for different $\kappa$ of the sample at different $\delta T$ values. As $\delta T$ increases, the change of $\delta R$ vs $\kappa$ increases, allowing for a more sensitive measurement. **Figure 4c** shows the line cut taken at $\kappa = 10$ W/m.K to show how relative resistivity change varies at different $\delta T$ values. One of the most important prospects of MBT is that the measurements can be performed at much lower $\delta T$ values than other methods. To better understand the limits imposed by intrinsic noise sources, we performed an in-depth error analysis at different $\delta T$ values. **Figure 4d** and **e** show how measured $\delta R$ varies for the material's thermal conductivity, at $\delta T = 0.1$ K and 1.0 K, respectively. Error from 0.1 to 1.0 K decreases by almost a factor of ten. Indeed, as shown in **Figure 4f**, the intrinsic error in measuring $\kappa$ decreases exponentially with the increasing $\delta T$. Above $\delta T = 0.3$ K, the error becomes $\sigma_\kappa = \pm 1.4$ W/m.K for $\kappa = 10$ W/m.K. Change of resistance for different thermal conductivity values are provided in the Supporting Information.

**Non-Fourier Heat Transport Measurements with MBT**
One of the prospects of MBT is exploring non-Fourier heat transport regimes where heat is transported in a non-diffusive manner. Particularly, this regime is relevant to MBT-based studies thanks to its controllable heater size and suspended crystal size. To test the applicability of MBT in non-Fourier heat transport regimes such as the hydrodynamic regime, we performed FEA simulations in the steady state. We modeled the heat transport in the hydrodynamic regime using the Guyer–Krumhansl equation[35]:

$$\vec{q} + \kappa \vec{\nabla} T = l^2 \nabla^2 \vec{q} \quad \text{(Eq. 4)}$$

Here, $\vec{q}$ is the heat flux density, and $l$ is the phonon mean free path. For the sake of consistency with the previous results, we implemented the material parameters for $Ce_3PtIn_{11}$ in our non-Fourier thermal conductivity simulations. We would like to note that there is no experimental evidence that suggests hydrodynamic heat transport in $Ce_3PtIn_{11}$.

Details regarding the COMSOL simulations are provided in the Supporting Information. The simulations are performed at a constant tip temperature, with $\delta T = 4$ K and 3.1 K, at 3 K sample temperature. Like the previous simulations, since there is no thermal conductivity figure for $Ce_3PtIn_{11}$, we assumed $\kappa = 10$ W/m.K for $l$-dependent simulations. **Figure 5a-b** shows thermal maps with $l = 0.01$ µm and $l = 2$ µm as a comparison, respectively. When $l$ is very short, the Guyer-Krumhansl equation produces a thermal distribution over the suspended region of the sample, which is similar to the Fourier regime. However, for a longer $l$, the hot spot becomes localized, consistent with the ballistic nature of the heat transport. **Figure 5c** shows thermal profiles along the sample region for various $l$ values. There is a clear distinction between the ballistic and diffusive regimes. To understand the phonon mean free path measurement sensitivity of MBT for a material with known thermal conductivity, we simulated the resistance change versus the phonon mean free path, as shown in **Figure 5d**. Even though for the diffusive and ballistic transport regimes, MBT fails to provide a sensitive measurement of the phonon mean free path in the cross-over from diffusive to ballistic, namely when $l$ is comparable to the diameter of the suspended crystal section, MBT can be used in measuring $l$.

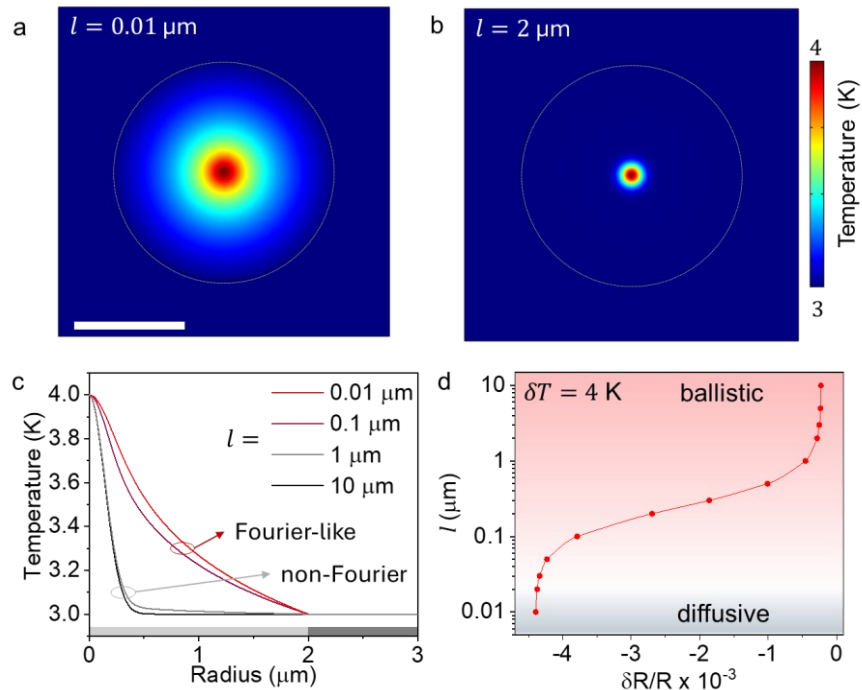

**Figure 5 | Non-Fourier simulations. a** Temperature map of a 30 nm thick $Ce_3PtIn_{11}$ sample simulated using Guyer-Krumhansl equation, when the AFM tip is at the center of the suspended region for kappa, $\kappa = 10$ W/m.K and $l = 0.01$ μm. The White dashed circle has a diameter of 4 μm and represents the suspended region of the sample. The scale bar represents 2 μm. **b** $l = 2$ μm shows a different thermal distribution over the suspended region. $\delta T = 4$ K. **c** Temperature line profiles from the center of the suspended crystal to the edge of the simulation are given for various $l$ values. For short phonon mean free path lengths, simulation yields Fourier-like, diffusive heat transport, while for longer $l$ values profile complies with ballistic transport. Light gray and dark gray shaded regions show the suspended and supported parts of the crystal, respectively. **d** $l$ vs. $\delta R/R$ is shown for $\kappa = 10$ W/m.K with $\delta T = 4$ K. For diffusive and ballistic transport regimes, the measurement sensitivity of $l$ decreases.

**Limitations of the MBT**

Despite the widespread proposed applicability of the MBT on various materials, MBT loses resolution in certain cases. The first limitation is when the signal-to-noise ratio becomes immeasurably small at above ~GΩ resistance levels. The reason is the limited modulation frequency due to the slow RC time constant of the device. However, in principle, Joule heating modulation to the tip can be used at much lower frequencies than the AFM modulation frequency (~10 kHz) to improve the signal-to-noise ratio with a compromise of DC temperature increase.

Another limitation of MBT that should be considered is the presence of zero-slope regions in temperature-dependent resistivity. This either means resistivity does not respond to temperature changes or resistivity has inflection point(s), as discussed in detail in the Supporting Information. In the former case, the method will not work until a meaningful temperature-dependent resistance change is registered. However, it should be noted that such resistivity plateaus in temperature are very rare, and other methods can be employed in conjunction with MBT to explore such regimes. In the latter case, inflection points can be handled in a way that the input power results in a temperature change sufficiently low not to cross the inflection point. As a result, inflection points do not pose a fundamental limit to MBT but rather a technical obstacle to be aware of during the experiments. Similarly, superconducting materials below the critical point cannot be studied. Finally, MBT is a steady-state method and cannot provide insight into ultra-fast heat transport in its current design.

**Conclusions**

In this manuscript, we proposed and studied the feasibility of a new thermal conductivity measurement method, MBT, based on the use of bolometric response of materials as a means of thermometry. MBT is based on heat exchange via a hot AFM tip. As a result, MBT has the potential to be a versatile tool that can work on a wide range of materials with a resolution reaching down to ~20 nm. Measurement sensitivity is comparable to cutting-edge methods like SQUID on tip and can be applied at a wide range of temperatures as long as electrical resistivity responds to thermal change induced on the sample. MBT can offer a systematic method to investigate thermal transport in low-dimensional and nanoscale materials, and can be used to study fundamental open questions in thermal transport related phenomena and quantum systems. An experimental demonstration of the proposed feasibility of MBT lies as an open challenge.

**Author Contributions**

TSK conceded the idea performed the Fourier regime simulations, wrote the manuscript. BK performed the non-Fourier regime simulations and commented on the manuscript. All authors discussed the results.

**Acknowledgements**

TSK acknowledges funding from TUBITAK under Grant#123F129.

# Supporting Information: An extensive thermal conductivity measurement method based on atomic force microscopy

T. Serkan Kasırga[1,2], Berke Köker[3]
[1] Bilkent University UNAM – National Nanotechnology Research Center, Bilkent, Ankara 06800 Türkiye
[2] Institute of Material Science and Nanotechnology, Bilkent University, Ankara 06800 Türkiye
[3] Department of Physics, Bilkent University, Bilkent, Ankara 06800 Tükiye

## 1. Eliminating the effect of contact resistance

In two-terminal bolometric thermometry, we measure the current change induced by the bolometric effect, $I_{bolo}$, *under constant voltage bias*, $V$, and correlate it to the resistance change: $\delta R = \frac{V}{I_0 + I_{bolo}} - \frac{V}{I_0}$. Here, $I_0$ is the "cold" current. The equation can be simplified as $\delta R = -\frac{R^2}{V}\frac{I_{bolo}}{1+\frac{I_{bolo}}{I_0}}$ and $\delta R \cong -\frac{R^2}{V}I_{bolo}$ when $I_{bolo} \ll I_0$. $V$ can be increased to enhance $I_{bolo}$ signal. One challenge in constant voltage measurements is the contact resistance, $R_c$. $R_c$ can be detrimental in two-terminal measurements as the effective bias[1] on the device, $V_{eff}$ is proportional to the ratio of the resistance of the sample to the total resistance: $V_{eff} = V\frac{R}{R+R_c}$. As a result, for large contact resistance samples, $I_{bolo}$ can significantly decrease. One way around is performing bolometric voltage measurement *under constant current* in four or three terminal configurations. In voltage mode, instead of constant voltage, constant current, $I_0$ is supplied to the sample. Then, the voltage change due to the bolometric effect, $V_{bolo}$ is measured. The resistance change can be written as: $\delta R = \frac{V+V_{bolo}}{I_0} - \frac{V}{I_0} = \frac{V_{bolo}}{I_0}$. Here, the sensitivity can be increased by increasing the applied current. Any zero-bias effect can be subtracted by performing a zero-current measurement.

## 2. Quantifying heat exchange by using SThM

When an SThM tip is used, heat exchange can be quantified, instead of guessing from the environmental parameters. The SThM tip is heated via Joule heating by the resistive elements fabricated into the tip. The power from Joule heating, $P_{JC}$ will be dissipated to the sample, $\dot{Q}_{TS}$, and to the environment, $\dot{Q}_{TE}$ (cantilever, exchange gas, etc.). We can define two thermal conductance paths for the tip: tip to sample $G_{TS}$ and tip to cantilever (or to the environment) $G_{TE}$. We are interested in finding the heat flux from tip to sample, $\dot{Q}_{TS}$. For measuring $\dot{Q}_{TE}$ cantilever can simply be raised to a height of no interaction with the substrate ($\dot{Q}_{TS} = 0$), and the heating current can be set to maintain constant tip temperature[2]. $P_{JC}$ can simply extracted from the resistance of the cantilever and the applied current. To improve the accuracy, we can also employ a method based on Menges et al.[3,4]. The method relies on modulating the sample temperature by Joule heating through a small ac voltage or current at frequency $f$ applied to the sample. I would like to emphasize that MBT normally operates at no Joule heating regime; however, for calibration of the tip, the following quantification must be performed once for each material. The Peltier response of the sample can also be measured and incorporated in the response, but since we only need $\dot{Q}_{TS}$ for calibration, we will not consider the Peltier term (depends on $f$). Let's assume a sinusoidal current is applied, $I = I_0 \sin(\omega t)$, which will lead to sample temperature modulation:

$$T_{sample} \cong T_0 + T_{sample,2f}(1 + \sin(2\omega t))$$

assuming thermal time of the sample is sufficiently fast to create a non-varying constant signal due to Joule heating. Here, $T_0$ is ambient temperature. The variation of the sample temperature leads to variation of $\dot{Q}_{TS}$ as the heat created in the sample is transferred to the tip:

$$\dot{Q}_{TS} = \dot{Q}_{TS,DC} + \dot{Q}_{TS,2f}\sin(2\omega t)$$

Using a Wheatstone bridge, the temperature of the tip can be measured via resistance change, which is also modulated by the sample temperature modulation:

$$T_{tip} = T_{tip,DC} + T_{tip,2f}\sin(2\omega t)$$

and the following relation ca be derived:

$$T_{sample,2f} = \frac{T_{tip,2f}\dot{Q}_{TS,DC} - (T_{tip,DC} - T_0)\dot{Q}_{TS,2f}}{\dot{Q}_{TS,DC} - \dot{Q}_{TS,2f}}$$

Finally, the heat flux $\dot{Q}_{TS}$ needs to be related to the electrical power dissipated in the cantilever[2].

### 3. Resistance change at different tip temperatures and thermal conductivities

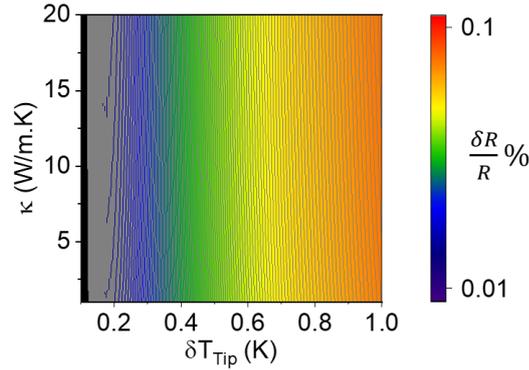

**Figure S1 | $\delta R/R$ change at different κ and $\delta T_{Tip}$ values. At $\delta T_{Tip}$ = 0.1 K, measurement remains below the noise level.**

### 4. Non-Fourier heat transport simulations

For the non-Fourier regime simulations, the same domain geometry, terminal voltage, and AFM tip parameters as the Fourier simulations were used. However, in the suspended part of the sample, the effect of convective heat transfer was neglected such that $\vec{\nabla} \cdot \vec{q} = U$, where $U$ is the heat entering the material from the AFM tip. Imposing this condition on the Guyer-Krumhansl equation, the following governing equation was obtained:

$$U + \kappa\nabla^2 T = l^2\nabla^2 U$$

Here, heat input was approximated as $U = \frac{Q}{t}e^{-r^2/r_0^2}$ with radius $r_o$ and $t$ same as those mentioned in Eq. 1 in the main text. Additionally, $Q$ was parametrically fitted to the temperature profile maximum such that for all $\kappa$ and $l$ values, the maximum of the sample temperature distribution matched the desired $T_{AFM}$ value.

### 5. MBT around resistivity plateaus and inflection points

As mentioned in the main text, one of the main limitations of the MBT is that in the absence of temperature-dependent resistivity change in a material measurement, sensitivity decreases or the method fails to provide a resistance change reading altogether. Another limiting case is the vicinity of inflection points in the temperature-dependent resistivity. Figure S1 shows the simulated response of a material in the vicinity of an inflection point. Depending on the input power, the system response changes from a negative response to no response. This is one of the limiting cases for the MBT.

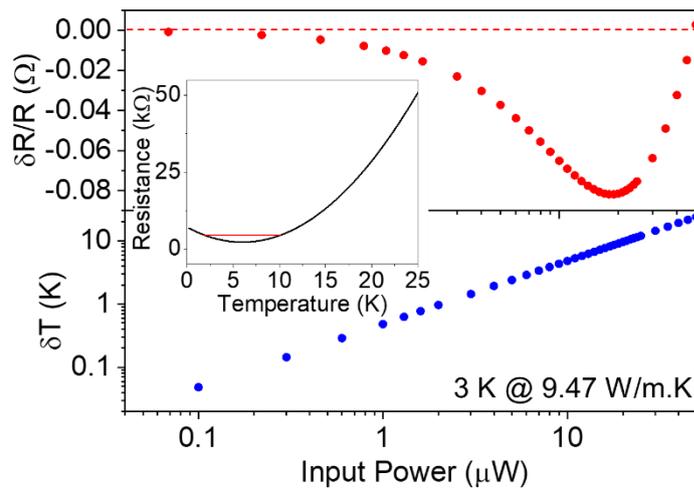

**Figure S2| Bolometric response around a resistance inflection point.** A thin metallic film exhibiting Kondo scattering with 9.47 Wm-1K-1 thermal conductivity is studied at 3 K using FEA. The inset shows the resistance vs. temperature graph. In the upper panel, δR/R vs. AFM input power shows that when the input power is large, the measured resistance change goes through a sign change. This shows that starting from 3 K, the resistivity of the sample first decreases, then increases due to the increasing local temperature. Such sign changes should be handled with care in MBT, as two very different values of thermal conductivity can be extracted. The lower panel shows the maximum temperature under the AFM tip.